\documentstyle[twocolumn,prl,aps,epsfig]{revtex}
\tighten
\newcommand{\beq}{\begin{eqnarray}}
\newcommand{\eeq}{\end{eqnarray}}
\renewcommand{\vec}[1]{{\mathbf{#1}}}
\begin{document}
\draft
\input epsf.sty

\title
{A Phase Glass is a Bose Metal: New Conducting State in Two Dimensions }
\author{Denis Dalidovich$^a$ and Philip Phillips$^b$}
\vspace{.05in}

%
\address
{$^a$National High Field Magnetic Laboratory,Florida State University,
Tallahassee, Florida\\
$^b$Loomis Laboratory of Physics,University of Illinois at Urbana-Champaign,
1100 W.Green St., Urbana, IL, 61801-3080}

%
\address{\mbox{ }}
\address{\parbox{14.5cm}{\rm \mbox{ }\mbox{ }
In the quantum rotor model with random exchange interactions
having a non-zero mean, three phases,
a 1) phase (Bose) glass, 2) superfluid,
and 3) Mott insulator, meet at a bi-critical point. 
We demonstrate that proximity to the bi-critical point
and the coupling between the energy landscape and the dissipative 
degrees of freedom of the phase glass lead to a
metallic state at $T=0$.
Consequently, the phase glass is unique in that it represents a concrete 
example of a metallic state that is mediated by disorder, even in 2D.
We propose that the experimentally observed metallic phase which intervenes
between the insulator and the superconductor in a wide range of thin films is
in actuality a phase glass.}}
\address{\mbox{ }}
\address{\mbox{ }}

\maketitle

There is now a preponderance of experimental 
evidence\cite{jaeger,mooij,yaz,mason} that
the disorder or magnetic field-induced destruction of 
superconductivity in a wide range of thin metal alloy films leads
first to a metallic state with a non-zero conductivity as $T\rightarrow 0$.
At sufficiently large values of the disorder or magnetic field,
a transition to a true insulating state obtains.  Within the standard
bosonic description\cite{mpaf,fwgf} of the insulator-superconductor transition (IST),
the onset of an intervening metallic state is problematic because
only two options are thought to exist for bosons: 1) localized 
in a Mott insulating state or 2) condensed in a superfluid.  
In the former, the conductivity vanishes whereas the latter exhibits 
resistanceless transport. Further,
including degrees of freedom which lie outside the bosonic or
phase-only models, for example electronic excitations, is of 
no help as electrons are localized in
2D. Indeed, while the onset of the insulator in homogeneously disordered
thin films is consistent\cite{valles,dynes,pmag} with the emergence 
of electronic excitations,
the intervening metallic and the subsequent superconducting states appear to 
be inherently bosonic in origin.

Consequently, recent theoretical 
effort\cite{dpglass,dpmet,dd} on the origin of the metallic
state has focused strictly on bosonic models.  Along these lines,
we have shown\cite{dpmet} that the standard Mott insulating phase in a clean
array of Josephson junctions has a non-zero conductivity as $T\rightarrow 0$.
This result arises from the non-commutativity\cite{ds}
 of the frequency and
temperature tending to zero limits of the conductivity in the vicinity of
a quantum critical point, with $\omega=0,T\rightarrow 0$ being the
experimentally relevant limit for the dc conductivity.
In the Mott insulator, quasiparticle excitations are gapped and obey
a Boltzmann distribution.  However, the collision time of such quasiparticles
grows exponentially with the gap\cite{dpmet}.  As the conductivity is a product of 
the collision time and the quasiparticle density, the conductivity
is necessarily finite in the limit $\omega<T$.  This type of Bose metal
is fragile\cite{dpmet}, however, and suppressed by dissipation and 
disorder.  In fact, in the presence of disorder, the nature of 
the superfluid-insulator transition changes dramatically.  For example,
several\cite{fwgf,sk} have argued that in the presence of on-site disorder, 
destruction of the superfluid may
(in the presence of incommensuration) obtain through an intervening 
phase with gapless excitations referred to as a Bose glass. 
In analogy with the Fermi glass, Fisher, et. al.\cite{fwgf} proposed
that the Bose glass is an insulator with variable-range hopping conductivity.

Nonetheless, we show explicitly that the glass phase which may 
interrupt the direct transition from a superfluid to a Mott insulator 
in the generally disordered case is 
in fact a metal that has a well-defined $T\rightarrow 0$ limit for the
conductivity.  As the thermal average of the superconducting
order parameter is non-zero but vanishes once averaged over disorder,
we refer to the glass as a phase glass.  We propose that 
the intervening metallic phase seen in the experiments is a phase glass.
Our proof that such a phase possesses a non-zero conductivity as 
$T\rightarrow 0$ constitutes the first demonstration
of a stable metallic state in 2D in the presence of disorder.

The starting point for our analysis is the charging model for an array
of superconducting islands
 \beq\label{HJ}
H=-E_C\sum_i\left(\frac{\partial}{\partial\theta_i}\right)^2-
\sum_{\langle i,j\rangle} J_{ij}\cos(\theta_i-\theta_j), 
\eeq
with random Josephson couplings $J_{ij}$ but fixed
on-site energies, $E_C$.   The phase of each island
is $\theta_i$.  Note that additional on-site disorder of the form
$iv_j\partial /\partial \theta_j$ results in the equivalent particle-hole symmetric
field theory provided that
the distribution of on-site energies has zero mean.  The non-zero
mean case is irrelevant here as this corresponds to a density-driven
IST\cite{sachbook}.
Hence, our conclusions apply to
the general disordered case.  To incorporate ordered phases, we assume 
that the Josephson energies are random and characterized by a Gaussian
distribution,
$P(J_{ij})=1/\sqrt{2\pi
J^2}\exp{-(J_{ij}-J_0)^2/2J^2}$,
with non-zero mean, $J_0$. The negative Josephson couplings included in this
distribution are essential to the physics of a disordered
superconductor\cite{sk}, particularly glassy ordering.  We have studied the 
non-zero mean problem extensively\cite{dpglass} and established explicitly
the existence of a bi-critical point in which three phases meet,
a Mott insulator, phase glass, and superconductor.
To distinguish between the phases, it is expedient
to introduce the set of variables ${\bf S}_i=(\cos\theta_i,
\sin\theta_i)$
which allows us to recast the interaction term
in the random Josephson Hamiltonian as a spin problem
with random magnetic interactions,
 $\sum_{\langle i,j\rangle}J_{ij}{\bf S}_i \cdot {\bf S}_j$.
Let $\langle ...\rangle$ and $[...]$ represent
averages over the thermal degrees of freedom
and over the disorder, respectively.  
In the superconductor not only $\langle S_{i\nu}\rangle$
but also $[\langle S_{i\nu}\rangle]$
acquire a non-zero value. In the phase (or spin) glass, however, 
$\langle S_{i\nu}\rangle\ne 0$ but $[\langle S_{i\nu}\rangle ]=0$,
whereas in the paramagnet or Mott
insulator, $\langle S_{i\nu}\rangle=0$. 

The Landau theory for this problem is easily obtained\cite{dpglass} using
replicas to average over the disorder and the identity 
$\ln [Z]=\lim_{n\rightarrow 0}([ Z^n]-1)/n$ to obtain the zero 
replica limit.  The quartic and quadratic spin-spin
interaction terms that arise from the disorder average can be decoupled
by introducing the auxiliary real fields,
\beq
 Q_{\mu\nu}^{ab}(\vec k,\vec k',\tau,\tau')=\langle
S_\mu^a(\vec k,\tau)S_\nu^b(\vec k',\tau')\rangle
\eeq
and $\Psi^a_\mu(\vec k,\tau)=\langle S^a_\mu(\vec k,\tau)\rangle$, 
respectively. The 
superscripts represent the replica indices.  A non-zero value of 
$\Psi^a_\mu(\vec k,\tau)$ implies phase ordering in
a charge $2e$ condensate.  Hence,
$\Psi^a_\mu$ couples to the charge degrees of freedom.
For quantum spin
glasses, it is the diagonal elements of the Q-matrix 
$D(\tau-\tau')=\lim_{n\rightarrow 0}\frac{1}{Mn}\langle
Q^{aa}_{\mu\mu}(\vec k,\vec k',\tau,\tau')\rangle$ in the limit that
$|\tau-\tau'|\rightarrow\infty$ that serve as the effective
Edwards-Anderson spin-glass order parameter\cite{sachdev,huse,bm}
within Landau theory.  The free energy per replica
\beq\label{fen}
&&{\cal F}[\Psi,Q]={\cal F}_{\rm SG}(Q)+
\sum_{a,\mu, k,\omega_n}(k^2+\omega_n^2+m^2)
|\Psi_\mu^a(\vec k,\omega_n)|^2 \nonumber\\
&&-\frac{1}{\kappa t}\int d^d x\int d\tau_1
d\tau_2\sum_{a,b,\mu, \nu}\Psi_\mu^a(x,\tau_1)
\Psi^b_\nu(x,\tau_2)
Q_{\mu\nu}^{ab}(x,\tau_1,\tau_2)\nonumber\\
&&+U\int d\tau\sum_{a,\mu}\left[\Psi_\mu^a(x,\tau)\Psi_\mu^a(x,\tau)\right]^2
\eeq
consists of a spin-glass part which is a third-order functional
of the $Q-$ matrices discussed  
previously\cite{dpglass,sachdev}, the $\Psi_\mu^a$ terms that describe 
the charge 2e condensate and the term
which couples the charge and glassy degrees of freedom. The parameters,
$\kappa$, $t$ and $U$ are the standard coupling constants in
a Landau theory and $m^2$ is the inverse correlation length.
In the disordered
phases, $\langle \Psi_\mu^a\rangle=0$; hence, in the glassy phase,
it is the fluctuations of the $\Psi_\mu^a$ field that survive.   Our previous
analysis\cite{dpglass} shows that
the cross term we have retained here is the most dominant
of the possible coupling terms near the bi-critical point.  

Our goal now is to calculate the charge transport in the glassy phase.
Near the spin-glass/superconductor boundary $m^2$ should be regarded as 
the smallest parameter. Hence, it is the fluctuations of $\Psi_\mu^a$ rather 
than those of $Q^{ab}$ that dominate.  Consequently,
we adopt the most general mean-field ansatz\cite{dpglass,sachdev} 
for the $Q-$ matrices
 \beq\label{qm}
Q_{\mu\nu}^{ab}(\vec k,\omega_1,\omega_2)&=&\beta(2\pi)^d\delta^d(k)
\delta_{\mu\nu}\left[
D(\omega_1)
\delta_{\omega_1+\omega_2,0}\delta_{ab}\right.\nonumber\\
&&\left.+\beta\delta_{\omega_1,0}\delta_{\omega_2,0}q^{ab}
\right].
 \eeq
The diagonal elements of the $Q$-matrices describe the excitation spectrum.
In the glassy phase, the spectrum is ungapped and given by
$D(\omega)=-|\omega|/\kappa$.  The linear dependence on $|\omega|$ arises
because the correlation function $Q^{aa}_{\mu\mu}(\tau)$ decays as
 $\tau^{-2}$\cite{sachdev,huse}.
 This dependence results in a fundamental change in the dynamical critical exponent from
$z=1$ to $z=2$ and the onset of overdamped dynamics (zero energy modes), thereby eliminating the $\delta(\omega)$ term from the conductivity.   In addition, without loss of generality, we work in 
the replica symmetric
case, $q^{ab}=q$ for all $a$ and $b$ as it was shown\cite{dpglass,sachdev} 
that replica symmetry breaking vanishes as $T\rightarrow 0$ and our 
emphasis is the low-temperature limit. Finally, because our focus is 
charge transport and the electromagnetic gauge couples only to 
the $\Psi^a_\mu$ field,
we retain only those terms in the free energy in which at least one of the 
$\Psi^a_\mu$ fields is present.    Substituting the $Q$-matrix ansatz 
(Eq. (\ref{qm})) into Eq. (\ref{fen}) and introducing a one-component complex
field $\psi^a=(\Psi^a_1,\Psi^a_2)$, we arrive at the following Gaussian 
theory,
\beq\label{fgauss}
{\cal F_{\rm gauss}}&=&\sum_{a,\vec k,\omega_n}
(k^2+\omega_n^2+\eta |\omega_n|+m^2)
|\psi^a(\vec k,\omega_n)|^2 \nonumber\\
&&-\beta q\sum_{a,b,\vec k,\omega_n} \delta_{\omega_n,0}
\psi^a(\vec k,\omega_n)[\psi^b(\vec k,\omega_n)]^{\ast},
\eeq
In the above action we introduced the effective dissipation 
$\eta=1/(\kappa^2 t)$ and rescaled $q\rightarrow q\kappa t$.
The associated Gaussian propagator is 
\beq\label{prop}
G^{(0)}_{ab}(\vec k,\omega_n)=G_0 (\vec k,\omega_n)\delta_{ab}+
\beta G_0^2(\vec k,\omega_n)q\delta_{\omega_n,0}
\eeq
in the $n\rightarrow 0$ 
limit\cite{zinjust} with
$G_0 (\vec k,\omega_n)=(k^2+\omega_n^2+\eta|\omega|+m^2)^{-1}$. 
The first term in
Eq. (\ref{prop}) is the standard Gaussian propagator in
the presence of Ohmic dissipation. The Ohmic dissipative term
in the free-energy arises from the diagonal elements of the 
$Q-$ matrices.  However, it is the $q-$dependent term
in the Gaussian free energy, the last term in Eq. (\ref{fgauss}), that is new
and changes fundamentally
the form of the propagator.  Because of the $\delta_{\omega_n,0}$ factor
in the second term in the free energy, the propagator now contains 
a frequency-independent part, $\beta G_0^2(\vec k,\omega_n=0) q$.  
In the free energy, this term couples different components of the replicas
and hence cannot be regrouped with the mass term, $m^2$. In  fact, this
term is a highly relevant perturbation in all dimensions. From simple 
tree-level scaling, 
$\psi=b^{(d+z+2)/2}\psi^{\prime}, k=k^{\prime}/b, \omega=
\omega^{\prime}/b^z$,
we find that the term proportional
to $q$ in the free energy rescales as $q^{\prime}=qb^{2+z}$.  The dynamical 
exponent $z$ is determined by the fact the scaling dimension of 
$\eta$ should remain
unchanged. This gives at the tree level
$z=2$ as proposed previously\cite{fwgf} for the superfluid/Bose glass
transition.  Hence, $q^\prime=qb^4$, implying that the coupling to the
energy landscape of the phase glass is strongly relevant
and ultimately responsible for the metallic phase. 

To see how this comes about, we use the generalization\cite{herbut} of the Kubo formula for 
the replicated action and write the conductivity to 
one-loop order per replica in the Gaussian approximation as
\beq
\sigma(i\omega_n)&=&\frac{2(e^*)^2}{n\hbar\omega_n}T\sum_{a,b,\omega_m}
\int \frac{d^2k}{(2\pi)^2} 
\left[G^{(0)}_{ab}(\vec k,\omega_m)\delta_{ab}\right.\nonumber\\
&&\left.-2 k_x^2 G_{ab}^{(0)}(\vec k, \omega_m)G_{ab}^{(0)}
(\vec k,\omega_m+\omega_n)\right].
\eeq
The conductivity contains
two types of terms.  All terms not proportional to $q$ have been 
evaluated previously\cite{flucon} and vanish as $T\rightarrow 0$. The terms
proportional to $q^2$ vanish in the limit $n\rightarrow 0$.  The only
terms remaining are proportional to $q$ and yield after an appropriate 
integration by parts
\beq
\sigma(i\omega_n)=\frac{8qe^{\ast 2}}{\hbar\omega_n}\int
\frac{d^2k}{(2\pi)^2}k_x^2G_0^2(\vec k,0)\left[ G_0(\vec k,0)-
G_0(\vec k,\omega_n)\right].\nonumber
\eeq
The momentum integrations are straightforward and yield
\beq
\sigma(\omega=0,T\rightarrow 0)=\frac{8e^2}{\hbar}\frac{q\eta}{2 m^4}
\eeq
a temperature-independent value for the conductivity as $T\rightarrow 0$.  
The dependence on
$q$ and $\eta$ implies that dissipation alone is insufficient
to generate a metallic state.  What seems to be the case
is that a bosonic excitation moving in a dissipative
environment in which many false minima exist does not localize because
it takes an exponentially long amount of time to find the ground state.
This is the physical mechanism that defeats localization in
a glassy phase.   
Further, the conductivity scales as $1/m^4$ and hence diverges as 
the superconducting
phase is approached. This is precisely what is seen experimentally
\cite{jaeger,mooij,yaz,mason}. 

That the conductivity plateaus in the phase glass regime does not appear
to have been anticipated previously.  We now appeal to much more general 
arguments to prove that the singular dependence of the
conductivity on $m^2$ as $T\rightarrow 0$ survives even
in the presence of the quartic interaction.
At the tree level, a dynamical exponent of $z=2$ renders 
the quartic interaction $U$ marginally irrelevant. However, considering
the last term in Eq. (\ref{fgauss}) on equal footing with $U$
in the one-loop renormalization group scheme, we reach the conclusion  
that the RG equations flow to strong coupling. The relevance of $q$ 
at all dimensions manifests itself also by the increasing singularity of 
relevant contributions from higher order diagrams in the perturbation 
series in $U$. We consider first the linear $U$ correction.
At this level, the self energy 
is given by a standard tad-pole diagram that arises from the couplings
in the average 
$\langle\psi^a\psi^{\ast b}\sum_c \psi^c\psi^{\ast c}\psi^c\psi^{\ast c}
\rangle$, yielding
$\Sigma=U\int_{\omega} \int_{\vec k} G^{(0)}_{aa}(\vec k,\omega_n)$. 
This diagram\cite{sachbook2} is regularizable only once the term 
$\omega_n^2$ is retained in the propagator. 
The first term in Eq. (\ref{prop}) leads at $T=0$ to a standard mass 
renormalization and innocuous logarithmic corrections, while
the last term is more singular, giving $\Sigma^{(1)}=Uq/(4\pi m^2)$.
A similar analysis can be undertaken to find the first-order correction to 
the $T=0$ conductivity. 
The relevant diagrams\cite{cha} can be readily generalized
for a two-replica propagator given by Eq. (\ref{prop}).
The straightforward evaluation of the contribution that does not vanish in 
the $T\rightarrow 0$ limit yields 
\beq
\sigma^{(1)}(\omega=0) =\frac{3e^2}{4h}\frac{U\eta q^2}{\pi m^6},
\eeq
 suggesting that each subsequent order in the interaction 
leads to a more singular contribution to the self-energy.  This points
to a scaling function of the form $\sigma \approx (e^2/\hbar)(\eta q/m^4)\Phi\left( q /m^2 \right)$ where $\Phi(y) \sim y^p$ for large $y$, which yields the critical behavior 
$\sigma \sim m^{-x}$ with $x = 4+2p$. The value of the exponent $p$ cannot 
inferred at any finite order in perturbation theory.

\begin{figure}
\begin{center}
\epsfig{file=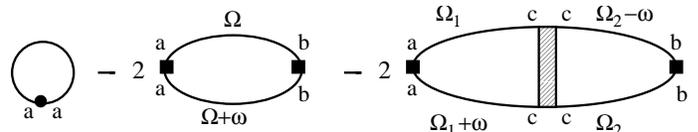, width=9cm}
\caption{Diagrammatic representation of the conductivity.  Each solid line
denotes the fully renormalized propagator, $G_{ab}$ 
(see Eq.(\protect\ref{gab})) while the shaded rectangle is
the vertex function. The letters $a,b,c$ represent the replica 
indices, and the internal momenta $\vec k_{i}$ are not shown 
for simplicity. }
\label{pgdiag}
\end{center}
\end{figure}
Nonetheless, we assume that all of the 
most singular diagrams can be resummed. A simple inspection of the 
perturbation series suggests that the fully renormalized propagator,
\beq\label{gab}
G_{ab}(\vec k,\omega_n)=\tilde{G}(\vec k,\omega_n)\delta_{ab}+
\beta q g(\vec k)\delta_{\omega_n,0},
\eeq
can be broken into replica diagonal and off-diagonal pieces.
Likewise, we define the self-energy associated with this propagator to be
$\Sigma_{ab}(\vec k,\omega_n)=\tilde{\Sigma}(\vec k,\omega_n)\delta_{ab}+
\beta q\theta(\vec k)\delta_{\omega_n,0}$ which contains formally 
all interaction terms. From the Dyson equation 
$G_{ab}=G^{(0)}_{ab}+G^{(0)}_{ac} \Sigma_{cd} G_{db}$,
in which the summation over the repeated indices is implied,
we have that 
$\tilde{G}(\vec k,\omega_n)=(G^{-1}_0(\vec k,\omega_n)-
\tilde{\Sigma}(\vec k,\omega_n))^{-1}$
and 
$g(\vec k)=[1+\theta(\vec k)]/
(G^{-1}_0(\vec k,0)-\tilde{\Sigma}(\vec k,0))^2$.
The renormalization of the interaction $U$ leads to the appearance of 
the corresponding vertex function\cite{zinjust} 
$\Gamma(\vec k_1,\vec k_2,\vec k;\Omega_1,\Omega_2,\omega_n)$ 
which is connected to the self-energy by means of the standard Dyson 
equation. This vertex function enters the general expression for the
conductivity represented diagrammatically 
in Fig. (\ref{pgdiag}). 
\noindent Each solid line represents the renormalized
propagator, $G_{ab}$, while the shaded region denotes the vertex function,
$\Gamma(\vec k_1,\vec k_2,\vec k=0;\Omega_1,\Omega_2,\omega_n)$.
We are interested here only in the static $T=0$ conductivity.
Once the first term in this diagrammatic expansion
is integrated by parts, use of the standard Ward identity leads
immediately to a cancellation of all diagrammatic contributions
to $\sigma$ in which the external frequency vanishes. 
As a consequence, we obtain the 
leading contribution to the 
conductivity in the limit that $\omega=0,T\rightarrow 0$ simply from a
Taylor expansion around $\omega=0$.  Using Eq. (\ref{gab}) for the 
renormalized Green function, we obtain the exact expression,
\beq\label{t0}
\sigma &=&\frac{32\pi e^2}{h}
q\left[ 2\int\frac{d^2k}{(2\pi)^2} k_x^2 g(\vec k)
\left(-\frac{\partial \tilde{G}(\vec k,|\Omega|)}{\partial|\Omega|}
|_{\Omega=0}\right)\right.\nonumber\\
&&\left.-q\int\frac{d^2k_1d^2k_2}{(2\pi)^4}k_{1x}k_{2x} g
(\vec k_1) g(\vec k_2)\tilde{G}(\vec k_1,0)\right.\nonumber\\
&&\left.\left[2\frac{\partial\tilde{G}(\vec k_2,|\Omega|)}{\partial|\Omega|}
|_{\Omega=0}\tilde{\Gamma}(\vec k_1,\vec k_2,0)\right.\right.\nonumber\\
&&\left.\left.+\tilde{G}(\vec k_2,0)\frac{\partial\tilde{\Gamma}}
{\partial|\Omega|}
(\vec k_1,\vec k_2,|\Omega|)\|_{\Omega=0}\right] \right],
\eeq
for the temperature-independent part of the conductivity.  Here
$\tilde{\Gamma}(\vec k_1,\vec k_2,|\omega_n|)=
\Gamma(\vec k_1,\vec k_2,0;-\omega_n,\omega_n,\omega_n)+
\Gamma(\vec k_1,\vec k_2,0;0,0,\omega_n)$, and it is taken into account
that the frequency dependence of all functions enters through 
$|\omega_n|$ due to the full particle-hole symmetry. In deriving Eq. 
(\ref{t0}), we assumed that, 1) the infinite perturbation series in
$U$ is resumable in principle, and 2) that all propagators and the 
vertex function are analytic in $|\omega_n|$. The latter assumption 
seems reasonable, because the most singular contributions come from 
diagrams that do not contain frequencies at all.    

We have demonstrated here that the sluggish phase dynamics in a
phase glass leads ultimately to a metallic
state in $d=2$ for bosonic excitations.  The strong divergence 
of the resultant conductivity on $m^2$ is consistent
with the experiments that have observed a distinct plateauing of 
the resistivity at low temperatures which increases in 
magnitude\cite{jaeger} as the distance from the true superconducting 
phase is increased. The metallic $T=0$ behavior obtains as a result of the 
coupling between the dissipative environment and the energy landscape
of the phase glass.  Further, since the dissipation inherent in
a phase glass is independent of temperature, external dissipation arising
from phonons is irrelevant as such coupling vanishes as $T\rightarrow 0$. 
Consequently, the metallic phase we have found here is robust to disorder
and phonon scattering and in fact constitutes the first explicit demonstration
of a metallic state in $d=2$.
Nonetheless, the theory presented here does not address 
the issue of whether or not the destruction of the superconducting phase occurs 
directly into a conventional insulator or a glassy phase. 
However, we suggest that only in the second scenario is 
the destruction of a 2D superconductor in the absence of a magnetic field 
consistent with the robustness of the metallic phase with 
respect to increasing disorder. 
We propose that aging and noise measurements as well as experiments
sensitive to trapped flux should be performed in the intervening
metallic regime to explore the glassy scenario suggested here. Clearly
a promising extension of this work would be the fermionic case.

\acknowledgements We thank S. Kivelson for pointing out Ref.
14, S. Sachdev and V. ${\rm Dobrosavljevi{\acute c}}$ for a careful reading of our paper, Antonio
Castro Neto for encouragement, and 
P. Goldbart, E. Fradkin, B. Halperin and D. S. Fisher for
useful discussions on disorder.
This work was partially funded by donors of the ACS petroleum research fund.

\end{document}